\documentclass[11pt]{article}

\usepackage{amstext,amsgen,latexsym,amsmath}
\usepackage{amstext,amssymb,amsfonts,latexsym}
\usepackage{theorem} \usepackage{pifont}\usepackage[dvips]{graphics,epsfig}

\def\Tr{{\rm Tr}}

 \newcommand{\bs}{\bigskip} 
 \newcommand{\n}{\noindent} 
 \newcommand{\hs}[1]{\hspace*{ #1 mm}}

 \def\bbox{\vrule height6pt width6pt depth1pt}

\theoremstyle{plain}
\theoremheaderfont{\bfseries}
 \newtheorem{theorem}{Theorem}[section] \newtheorem{lemma}[theorem]{Lemma}
 \newtheorem{proposition}[theorem]{Proposition}

 \newenvironment{proof}{\par \noindent
            {\bf Proof. \hs{2}}}{\hfill$\Box$ \vspace*{3mm}}
 
 \newenvironment{proofof}[1]{\vspace*{5mm} \par \noindent
         {\bf Proof of #1.\hs{2}}}{\hfill$\Box$ \vspace*{3mm}}

\newcommand{\ignore}[1]{}
\begin{document}
\pagestyle{plain}


\begin{center}
{\Large {\bf Quantum Network Coding for General Graphs}}
\bs\\

\begin{center}
{\sc Kazuo Iwama}$^1$\footnote{Supported in part by Scientific Research Grant, Ministry of Japan, 16092101.} 
\hspace{5mm} {\sc Harumichi Nishimura}$^2$\footnote{Supported in part by Scientifis Research Grant, 
Ministry of Japan, 18244210.}
\hspace{5mm} {\sc Rudy Raymond}$^3$ \hspace{5mm} {\sc Shigeru Yamashita}$^4$\footnote{Supported in part by Scientific Research Grant, Ministry of Japan, 16092218.} 
\end{center}

$^1${School of Informatics, Kyoto University}

{\tt iwama@kuis.kyoto-u.ac.jp} 

$^2${School of Science, Osaka Prefecture University}

{\tt hnishimura@mi.s.osakafu-u.ac.jp}

$^3${Tokyo Research Laboratory, IBM Japan}

{\tt raymond@jp.ibm.com}

$^4${Graduate School of Information Science, Nara Institute of Science and Technology} 

{\tt ger@is.naist.jp}. 
\end{center}
\bs

\n{\bf Abstract.}\hs{1} 
Network coding is often explained by using a small network
model called Butterfly.  In this network, there are two flow paths,
$s_1$ to $t_1$ and $s_2$ to $t_2$, which share a single bottleneck
channel of capacity one.  So, if we consider conventional flow (of
liquid, for instance), then the total amount of flow must be at most
one in total, say 1/2 for each path.  However, if we consider
information flow, then we can send two bits (one for each path) at the
same time by exploiting two side links, which are of no use for the
liquid-type flow, and encoding/decoding operations at each node. 
This is known as {\em network coding} and has been quite popular 
since its introduction by Ahlswede, Cai, Li and Yeung in 2000. 
In QIP 2006, Hayashi et al showed that {\em quantum} network coding is
possible for Butterfly, namely we can send two {\em qubits} simultaneously 
with keeping their fidelity strictly greater than $1/2$.  

In this paper, we show that the result can be extended to a 
large class of general graphs by using a completely different 
approach. The underlying technique is a new cloning method called
{\em entanglement-free cloning} which does not produce any
entanglement at all. This seems interesting on its own and 
to show its possibility is an even more important purpose of this paper.
Combining this new cloning with approximation of general quantum
states by a small number of fixed ones, we can design a quantum
network coding protocol which ``simulates'' its classical counterpart for
the same graph.

\section{Introduction} 
In some cases, digital information flow can be done much more efficiently than 
conventional (say, liquid) flow. For example, consider the Butterfly network in Fig.\ 1 having directed
links of capacity one and two source-sink pairs $s_1$ to
$t_1$ and $s_2$ to $t_2$. Apparently, both paths have to go through
the single link from $s_0$ to $t_0$ (the two side links from $s_1$ to
$t_2$ and $s_2$ to $t_1$ are of no use at all) and hence the total
amount of flow is bounded by one, say $1/2$ for each pair. For
information flow, however, we can send two bits, $x$ and $y$,
simultaneously by using the protocol in Fig.\ 2. Such a protocol, by
which we can effectively achieve larger channel capacity than can be
achieved by simple routing, has been referred to as {\it network coding} since its introduction in \cite{ACLY00}.

In \cite{HINRY06-qip}, the authors proved that {\em quantum} network coding (QNC) is
possible for the same Butterfly network, namely, we can send two {\em
qubits} simultaneously with keeping their fidelity strictly greater
than $1/2$. They also showed that QNC is no longer possible or the
worst-case fidelity becomes 1/2 or less, if we remove the two side
links. Classical network coding (CNC) for this reduced network is also
impossible. Thus, CNC and QNC are closely related in Butterfly and we
are naturally interested in a similar relation for general graphs. A
typical question to this end is whether QNC is possible for the graph
class $\mathcal{G}(\mathbb{F}_2)$ (including Butterfly and many
others, see e.g., \cite{AHJKL06,HKL04}) which allows CNC by using linear operations
over $\mathbb{F}_2$ at each node.  This has been an obvious open question since \cite{HINRY06-qip}.

The crucial difference between CNC and QNC happens at a node with two
or more outgoing edges, where we need some kind of ``copy'' operation.
($s_1$, $s_2$ and $t_0$ in Fig. 1 are such nodes.)  In the case of
CNC, nothing is hard; just a usual copy operation is optimal.  In the
case of QNC, we first encounter the famous no-cloning theorem \cite{WZ82}.
This difficulty might be bypassed by using the approximate cloning
by Bu\v{z}ek and Hillery \cite{BH96} with a sacrifice of fidelity, 
but then arises another much more serious problem; entanglement between cloned
states.  Note that entanglement extends to the whole graph.  
In \cite{HINRY06-qip}, our analysis needed to explicitly observe the total state on the seven
edges of the Butterfly network.  It is very unlikely that we can
stay on the same approach for general graphs.

\begin{figure}[b]
\begin{minipage}{0.5\hsize}\begin{center}
\includegraphics*[width=4.5cm]{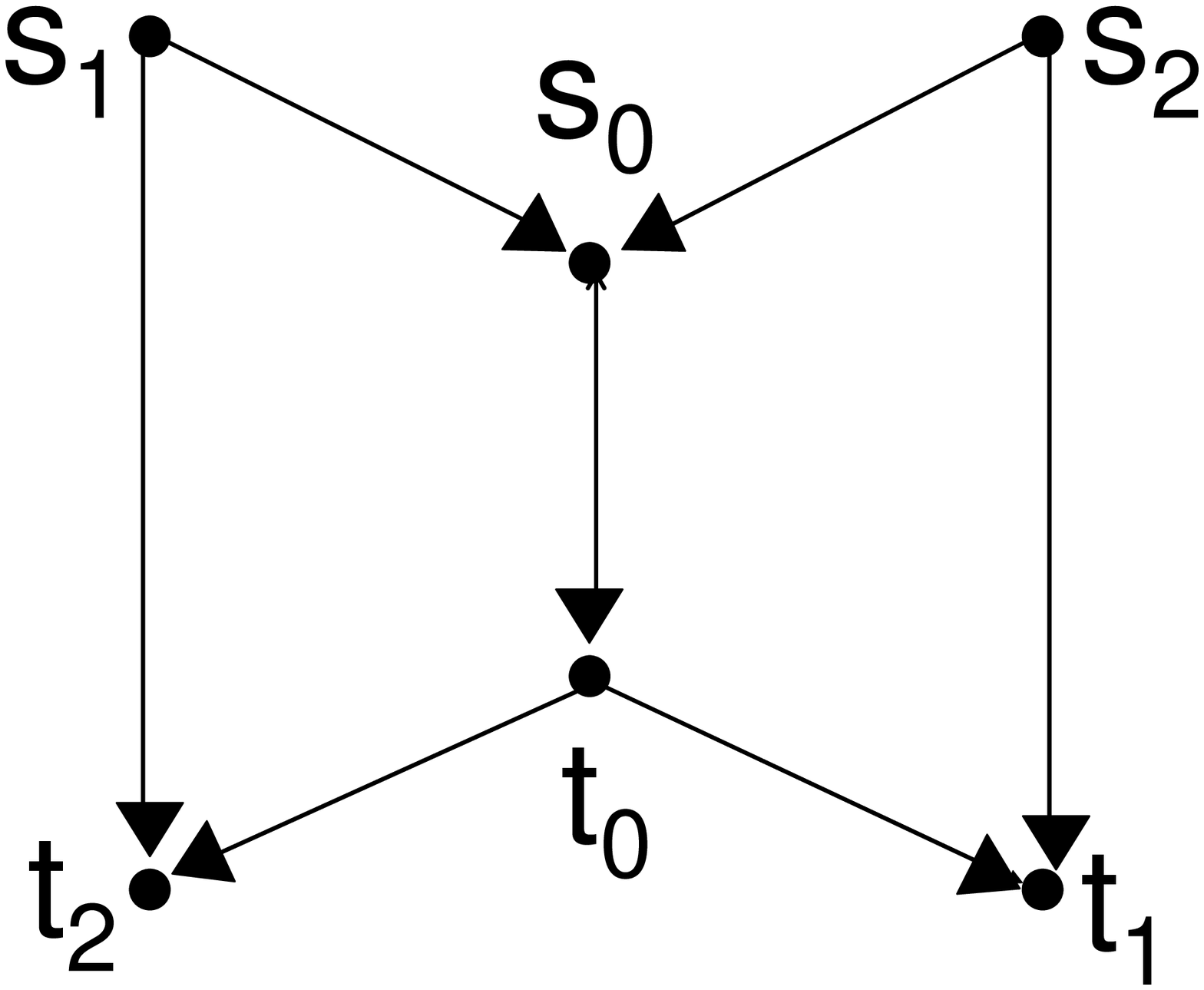}

Figure 1
\end{center}\end{minipage}
\begin{minipage}{0.33\hsize}\begin{center}
\includegraphics*[width=5cm]{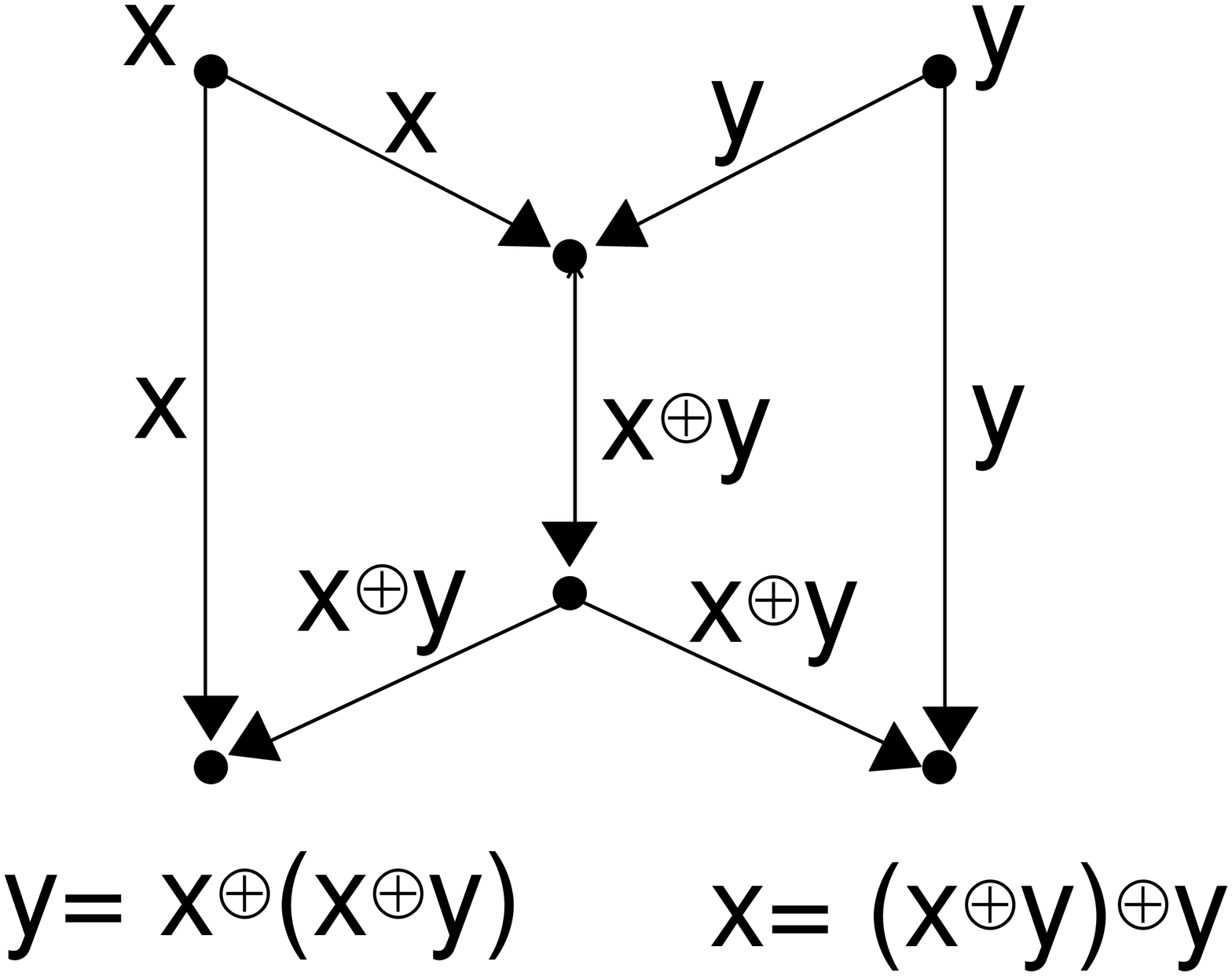}

Figure 2

\end{center}\end{minipage}
\end{figure}

{\bf Our Contribution.} In this paper, we give a positive answer to
the open question even for the much larger graph class $\mathcal{G}_4$: the graph class which allows some nonlinear
operations over a size-four alphabet to achieve CNC. 
$\mathcal{G}_4$ includes the above $\mathcal{G}(\mathbb{F}_2)$ 
and also many other graphs for which linear operations are not enough for CNC 
(see the next section for details).  For a given $G$ in $\mathcal{G}_4$ 
and a CNC protocol which sends any one letter in the alphabet correctly from
each source to sink, we can design a QNC protocol which sends an
arbitrary qubit similarly with fidelity $>1/2$.

Our key technique is a new cloning method called {\em
entanglement-free cloning}, which we believe is interesting in its own
right.  By using this cloning at each branching node, we no longer
need to observe the entire state of $G$ explicitly but it is enough
to calculate the quantum state at each node {\em independently}.
Combining this with approximation of quantum states by four fixed
ones, we can design a QNC protocol which ``simulates'' the given CNC protocol.

{\bf Related Work.} \cite{HINRY06-qip} inspired several studies on quantum network coding. 
Shi and Soljanin \cite{SS06} investigated the quantum network coding for the so-called
multi-cast problem where the graph has only one source node.
Leung, Oppenheim and Winter \cite{LOW06} discussed an asymptotic
limit of quantum network coding for graphs of low depth, including the Butterfly network. 
\cite{HINRY06} showed the impossibility of the $(4,1)$-quantum random access coding and 
its relation to quantum network coding

Quantum cloning has been one of the most popular topics. Its studies
are divided into the two types; the universal cloning and the
state-dependent cloning. The universal cloning, initiated by Bu\v{z}ek and
Hillery \cite{BH96}, and its successors (say, \cite{Bru98,GM97,Wer98}) 
produce approximated copies of any quantum state equally well. On the other hand, the input
of the state-dependent cloning is restricted to a fixed set of quantum
states, which has two different directions (and their hybrid such as
\cite{CB99}). The first one, introduced in \cite{HB97} and
further studied in \cite{Bru98}, is to seek a better quality by
limiting the input of the universal cloning into fixed states.
The goal of the second approach is to exactly clone quantum states in a probabilistic manner, 
so this is called the probabilistic cloning. The probabilistic cloning was proposed by Duan
and Guo \cite{DG98} and seems most related to ours (see Sec. \ref{sec:previous-work} for further details).

\section{Classical and Quantum Network Coding}\label{sec:cnc-qnc}

\subsection{Classical Network Coding}
For (classical) network coding, an instance is given as a directed acyclic graph $G=(V,E)$, 
a set $S=\{s_1,\ldots,s_n\}\subseteq V$ of $n$ {\em source nodes}, 
a set $T=\{t_1,\ldots,t_m\}\subseteq V$ of $m$ {\em sink nodes} 
and a {\em source-sink requirement} given by a mapping 
$\sigma:\{1,\ldots,m\}\rightarrow \{1,\ldots,n\}$, meaning that 
an input value on node $s_i$ should be sent to node(s) $t_j$ such that $\sigma(j)=i$. 
(Precisely, this does not contain the case that the sink requires multiple sources, 
but it is easy to adapt our result to that case.) 
Each link $e\in E$ has a unit capacity, i.e., it can transmit a single letter in a fixed alphabet $\Sigma$. 
A {\em network code} (or a {\em protocol}) for $G$, denoted by $P_C(G)$, 
is defined by $l$ functions (called {\em operations}) $f_{v,j}$: 
$\Sigma^k\rightarrow\Sigma$, $j=1,2,\ldots,l$, for each vertex $v\in V$ 
with indegree $k$ and outdegree $l$. We say that classical network coding (CNC) {\em is possible} 
if there is a protocol such that input values $(x_1,\ldots,x_n)$ 
given to the source nodes $S=\{s_1,\ldots,s_n\}$ imply the output values $(y_1,\ldots,y_m)$ 
on the sink nodes $T=\{t_1,\ldots,t_m\}$ such that $y_{j}=x_{\sigma(j)}$. 

Li, Yeung and Cai showed in \cite{LYC03} that if $G$ has only one source, {\em linear} operations 
are enough, i.e., if CNC is possible for such a graph, it is possible only by using linear operations 
over a finite (but maybe large) field. However, this is not the case for graphs with 
two or more sources: The first example, known as the Koetter's example, was given in \cite{MEHK03} 
where it is shown that the graph does not have a linear CNC 
even if its alphabet size is arbitrarily large, but does have a CNC 
if ``vector'' linear operations over an alphabet of size four (actually $\mathbb{F}_2^2$) are allowed. 
Very recently another example appeared in \cite{DFZ05}, which does not have a vector linear CNC 
over any alphabet, but has a CNC if we allow some non-linear operations over a size-four alphabet.  

In this paper we consider the following operations over a size-four alphabet which covers 
both \cite{MEHK03} and \cite{DFZ05}: Let $\Sigma_4=\{00,01,10,11\}$ and let $v$ be a node 
of indegree $m$. Then if the values of $m$ incoming edges are $X_1,\ldots,X_m\in\Sigma_4$, 
then the output of each outgoing edge can be written as $\sum_{i=1}^m h_i(X_i)$. 
Here, the summation is taken under the additive groups $\mathbb{Z}_4$ or 
$\mathbb{Z}_2\oplus \mathbb{Z}_2$ (note that additive groups over $\Sigma_4$ includes 
only $\mathbb{Z}_4$ and $\mathbb{Z}_2\oplus \mathbb{Z}_2$), and $h_i$ ($i=1,\ldots,m$) 
is any constant, one-to-one or two-to-one mapping over $\Sigma_4$. 
If $G$ has a CNC under these operations, we say that $G$ is in the graph class ${\cal G}_4$. 
As mentioned before, ${\cal G}_4$ includes both examples in \cite{MEHK03} and \cite{DFZ05}, 
and of course all the graphs (including Butterfly) for which CNC is possible 
by linear operations of size two and four. 

\subsection{Quantum Network Coding}

In quantum network coding (QNC), we suppose that each link of the graph $G$ is a quantum channel of capacity one, 
i.e., it can transmit a single quantum bit. At each node, any trace-preserving completely-positive (TP-CP) map 
is allowed. A protocol $P_Q(G)$ is given as a set of these operations at each node. We say that 
{\em QNC is possible} for a given graph $G$ if there is a protocol $P_Q(G)$ which determines, 
for given input qubits $|\psi_1\rangle,\ldots,|\psi_n\rangle$ on the $n$ source nodes, 
outputs $\pmb{\rho}_1,\ldots,\pmb{\rho}_m$ on the sink nodes such that the fidelity 
between $\pmb{\rho}_j$ and $|\psi_{\sigma(j)}\rangle$ is greater than $1/2$. 
(Thus the inputs are pure qubits without entanglement and the output may be general mixed states. 
We often use bold fonts for density matrices for exposition.) 
Our main goal of this paper is to show that QNC is possible for any graph $G$ in ${\cal G}_4$, 
in other words, we can design a legitimate protocol $P_Q(G)$ 
from a given graph $G$ in ${\cal G}_4$ and its classical protocol $P_C(G)$.  

\section{Entanglement-Free Cloning}\label{sec:efc} 

\subsection{Basic Ideas of Designing $P_Q(G)$}

Our QNC is based on the following ideas: (i) If we carefully select a small number (say, four) 
of fixed quantum states, then any quantum state can be ``approximated'' by one of them. 
(ii) Therefore, if we can change a given state into its approximation at each source node, 
we can assume without loss of generality that each source node receives only one of these four states. 
Thus our task is to send it to its required sink node(s) as faithfully as possible. 
(iii) This can be obviously done by the following: Select a one-to-one mapping between the four quantum states 
and the four letters in $\Sigma_4$ and design a TP-CP map which simulates the classical operation for $\Sigma_4$ 
at each node. 

Now the question is how to design these quantum operations from its classical counterparts. 
It then turns out that it is not so hard to design ``main'' operations corresponding to 
$\sum_i h_i(X_i)$. The real hard part (the trivial part in the classical case) 
is to distribute this calculated state into two (or more) outgoing edges. 
The reason is, as one can expect easily, that entanglement is easily involved. 
Since the graph is arbitrarily complicated, there are a lot of different paths 
from one source to one sink which fork and join many times; it seems totally impossible 
to keep track of how the global entangled state is expanding to the entire graph. 
(In fact we need a lot of effort to cope with this problem even for the (very simple) 
Butterfly network \cite{HINRY06-qip}.) 

Our solution to this difficulty is {\em entanglement-free cloning (EFC)} that does not produce 
any entanglement between two outputs. Formally, EFC is defined as follows. 
A TP-CP map $f$ is an EFC for a set of quantum states ${\cal Q}=\{\pmb{\rho}_1,\ldots,\pmb{\rho}_m\}$ 
if there exist $p,q>0$ such that, for any $\pmb{\rho}\in {\cal Q}$, 
$f(\pmb{\rho})=(p\pmb{\rho}+(1-p)\frac{\pmb{I}}{2})\otimes (q\pmb{\rho}+(1-q)\frac{\pmb{I}}{2})$. 
If such a map exists, we say that ${\cal Q}$ admits an EFC. 

\subsection{Necessary Conditions for EFC}

Now our goal is to find a set of states which admits an EFC. We first prove the following necessary condition.

\begin{proposition}\label{efc-condition}
If a set ${\cal Q}=\{\pmb{\rho}_1,\ldots,\pmb{\rho}_m\}$ of quantum states admits an EFC, 
then $\pmb{\rho}_1,\ldots,\pmb{\rho}_m$ are linearly independent (on the vector space $M_2(\mathbb{C})$, 
the set of $2\times 2$ matrices on $\mathbb{C}$). 
\end{proposition}

\begin{proof}
Suppose for contradiction that $\pmb{\rho}_1,\ldots,\pmb{\rho}_m$ are not linealy independent. 
Namely, there exists an index $j$ such that $\pmb{\rho}_j = \sum_{i\neq j} c_i \pmb{\rho}_i$. 
Without loss of generality, we can assume $j = m$, that is, 
\begin{equation}\label{sum_st}
\pmb{\rho}_m = \sum_{i=1}^{m-1} c_i \pmb{\rho}_i.
\end{equation}
Notice that $\sum_{i=1}^{m-1} c_i = 1$ since $\Tr(\pmb{\rho}_m) = 1$, 
and that there are at least two non-zero $c_i$'s since any two states are linearly independent on $M_2(\mathbb{C})$. 
Moreover, we can assume that 
the states of ${\cal Q}\setminus\pmb{\rho}_m$ are linearly independent 
(otherwise, remove some elements from ${\cal Q}\setminus\pmb{\rho}_m$ until it becomes linearly independent). 

Suppose that ${\cal Q}$ admits an EFC. Then there is a TP-CP map ${\mathbf M}$ such that 
${\mathbf M}(\pmb{\rho}_i)=(p\pmb{\rho}_i+(1-p)\frac{\pmb{I}}{2})\otimes(q\pmb{\rho}_i
+(1-q)\frac{\pmb{I}}{2})$ where $p,q>0$. By the linearity of ${\mathbf M}$ and Eq.~(\ref{sum_st}) 
we have 
\begin{equation}\label{sum_efc}
{\mathbf M}(\pmb{\rho}_m) = \sum_{i=1}^{m-1} c_i {\mathbf M}(\pmb{\rho}_i),
\end{equation}
which implies the following relation. 
\begin{equation}\label{sum_efc2}
\left(p\pmb{\rho}_m + (1-p)\frac{\pmb{I}}{2}\right) \otimes \left(q\pmb{\rho}_m + (1-q)\frac{\pmb{I}}{2}\right) 
= 
\sum_{i=1}^{m-1}c_i \left(p\pmb{\rho}_i + (1-p)\frac{\pmb{I}}{2}\right) \otimes 
\left(q\pmb{\rho}_i + (1-q)\frac{\pmb{I}}{2}\right). 
\end{equation}
The left-hand side of Eq.(\ref{sum_efc2}) is rewritten as 
\[
pq \pmb{\rho}_m\otimes\pmb{\rho}_m+p(1-q)\pmb{\rho}_m\otimes\frac{\pmb{I}}{2}
+q(1-p)\frac{\pmb{I}}{2}\otimes\pmb{\rho}_m+(1-p)(1-q)\frac{\pmb{I}}{2}\otimes\frac{\pmb{I}}{2},
\]
and the right-hand as
\begin{align*}
&pq \sum_{i=1}^{m-1}c_i\pmb{\rho}_i\otimes\pmb{\rho}_i
+p(1-q)\sum_{i=1}^{m-1}c_i\pmb{\rho}_i\otimes\frac{\pmb{I}}{2}
+q(1-p)\sum_{i=1}^{m-1}c_i\frac{\pmb{I}}{2}\otimes\pmb{\rho}_i
+(1-p)(1-q)\sum_{i=1}^{m-1}c_i\frac{\pmb{I}}{2}\otimes\frac{\pmb{I}}{2}\\
&=pq \sum_{i=1}^{m-1}c_i\pmb{\rho}_i\otimes\pmb{\rho}_i
+p(1-q)\pmb{\rho}_m \otimes \frac{\pmb{I}}{2}+ q(1-p)\frac{\pmb{I}}{2}\otimes\pmb{\rho}_m
+(1-p)(1-q)\frac{\pmb{I}}{2}\otimes\frac{\pmb{I}}{2},
\end{align*}
where we used Eq.(\ref{sum_st}) and $\sum_{i=1}^{m-1} c_i=1$. 
Thus, by canceling the same terms we obtain $pq\pmb{\rho}_m\otimes
\pmb{\rho}_m=pq\sum_{i=1}^{m-1} c_i\pmb{\rho}_i\otimes\pmb{\rho}_i$, which implies  
$\pmb{\rho}_m\otimes\pmb{\rho}_m=\sum_{i=1}^{m-1} c_i\pmb{\rho}_i\otimes\pmb{\rho}_i$ since $pq\neq 0$. 
On the other hand, $\pmb{\rho}_m\otimes\pmb{\rho}_m=(\sum_{i=1}^{m-1} c_i\pmb{\rho}_i)^{\otimes 2}$ 
by Eq.(\ref{sum_st}) and hence we have 
\[
\sum_{i=1}^{m-1} c_i\pmb{\rho}_i\otimes\pmb{\rho}_i=\sum_{i,j=1}^{m-1} c_ic_j\pmb{\rho}_i\otimes\pmb{\rho}_j.
\]
Note that the states $\{\pmb{\rho}_i\otimes\pmb{\rho}_j\}_{i,j=1}^{m-1}$ are linearly independent 
since $\{\pmb{\rho}_i\}_{i=1}^{m-1}$ are linearly independent. Thus, for any $i,j\in\{1,2,\ldots,m-1\}$
\begin{equation}\label{eq1119}
c_i\cdot c_j =
\left\{\begin{array}{ll}
c_i &\ \mbox{if}\ j=i,\\ 
0   &\ \mbox{if}\ j\neq i. 
\end{array}
\right.
\end{equation}
Obviously $c_i=0$ or $1$ for any $i\in\{1,\ldots,m-1\}$. 
Since $\sum_{i=1}^{m-1}c_i=1$, there is only one index $i_0$ 
such that $c_{i_0}=1$ and $c_{j}=0$ for all other $j$. 
This contradicts the fact that there are at least two non-zero $c_i$'s. 
\end{proof}

Note that any two different states are linearly independent and thus satisfy the condition. 
In fact, we can show that any set of two states admits an EFC (see Appendix). 
Unfortunately, two states are not enough for our purpose since it is impossible to approximate 
an arbitrary quantum state 
with fidelity $>1/2$. For a set of four states, one can easily see that  
the BB84 states $\{|0\rangle,|1\rangle,|+\rangle,|-\rangle\}$, for instance, 
are not linearly independent and cannot be used, either. 

\subsection{EFC for Four States}

Our solution is to use what we call ``the tetra states'' defined by  
$|\chi(00)\rangle=\cos\tilde{\theta}|0\rangle+e^{\imath\pi/4}\sin\tilde{\theta}|1\rangle$, 
$|\chi(01)\rangle=\cos\tilde{\theta}|0\rangle+e^{-3\imath\pi/4}\sin\tilde{\theta}|1\rangle$, 
$|\chi(10)\rangle=\sin\tilde{\theta}|0\rangle+e^{-\imath\pi/4}\cos\tilde{\theta}|1\rangle$, 
$|\chi(11)\rangle=\sin\tilde{\theta}|0\rangle +e^{3\imath\pi/4}\cos\tilde{\theta}|1\rangle$ 
with $\cos^2\tilde{\theta}=\frac{1}{2}+\frac{\sqrt{3}}{6}$ (forming a tetrahedron in the Bloch sphere). 
It is straightforward to prove that $\{\pmb{\chi}(00),\pmb{\chi}(01),\pmb{\chi}(10),\pmb{\chi}(11)\}$ 
(where $\pmb{\chi}=|\chi\rangle\langle\chi|$) are linearly independent, but we still have to design 
an explicit map (protocol) for EFC. As shown below, our protocol fully depends on the {\em tetra measurement}, 
denoted by $TTR$, which is defined by the POVM (positive operator-valued measure) $\left\{\frac{1}{2}\pmb{\chi}(00),
\frac{1}{2}\pmb{\chi}(01),\frac{1}{2}\pmb{\chi}(10),\frac{1}{2}\pmb{\chi}(11)\right\}$. 
The following lemma is straightforward:

\begin{lemma}\label{ttr}
$TTR$ on $|\chi(z_1z_2)\rangle$ produces the two bits $z_1z_2$ with probability $1/2$, 
and the other three bits $z_1\bar{z}_2,\bar{z}_1z_2,\bar{z}_1\bar{z}_2$ with probability $1/6$. 
($\bar{z}$ is the negation of $z$.) Furthermore, the TP-CP map induced by $TTR$, 
$|\psi\rangle\mapsto \pmb{\chi}(TTR(|\psi\rangle))$, is {\em $1/3$-shrinking}, that is, 
$\pmb{\chi}(TTR(|\psi\rangle))=\frac{1}{3}|\psi\rangle\langle\psi|+\frac{2}{3}\frac{\pmb{I}}{2}$.  
\end{lemma}  

Now here is our protocol $EFC_\alpha$. The important point is that our cloning works not only 
for $\pmb{\chi}(X)$ where $X\in\Sigma_4$, but also for $\alpha\pmb{\chi}(X)+(1-\alpha)\frac{\pmb{I}}{2}$ 
if the value of $\alpha$ is known in advance.  

\

\noindent
{\bf Protocol $EFC_\alpha$.} Input: $\pmb{\rho}_\alpha=\alpha\pmb{\chi}+(1-\alpha)\frac{\pmb{I}}{2}$ 
where $\pmb{\chi}\in\{\pmb{\chi}(z_1z_2)\mid z_1z_2\in\Sigma_4\}$.

{\bf Step 1.} Apply the tetra measurement on $\pmb{\rho}_\alpha$, and obtain the two-bit measurement result 
$X\in\Sigma_4$. 

{\bf Step 2.} Produce the pairs of two bits $(Z_1,Z_2)$ from the measurement value $X$ 
according to the following probability distribution: $(X,X)$ with probability $p_1$; 
each of the forms $(X,Y)$ or $(Y,X)$ (6 patterns) with probability $p_2$ where $Y$ is a two bit different from $X$; 
each of the forms $(Y,Y')$ (6 patterns) with probability $p_3$ where $Y'$ is a two bit different from $X$ and $Y$; 
each of the forms $(Y,Y)$ (3 patterns) with probability $p_4$. 
(If $X=00$, for example, $(X,Y)=(00,01),(00,10)$, and $(00,11)$, $(Y,X)=(01,00),(10,00)$, and $(11,00)$, 
$(Y,Y')=(01,10),(01,11),(10,01),(10,11),(11,01)$, and $(11,10)$, and $(Y,Y)=(01,01),(10,10)$, and $(11,11)$.) 
Here, $p_1,p_2,p_3,p_4$ are positive numbers depending 
on $\alpha$ that are determined in the proof of Lemma \ref{EFC-lemma}. 
 
{\bf Step 3.} Send $|\chi(Z_1)\rangle$ and $|\chi(Z_2)\rangle$ to the two outgoing edges.

\begin{lemma}\label{EFC-lemma}
For any $\alpha>0$, $EFC_\alpha$ on input $\pmb{\rho}_\alpha$ produces the output 
$\left(\frac{\alpha}{9}\pmb{\chi}+\left(1-\frac{\alpha}{9}\right)\frac{\pmb{I}}{2}\right)^{\otimes 2}$. 
\end{lemma}

\begin{proof}
Notice that 
\begin{equation}\label{eq1111-0}
p_1+6p_2+6p_3+3p_4=1,
\end{equation}
which is the sum of probabilities. Let $\pmb{\chi}=\pmb{\chi}(z_1z_2)$ and 
suppose $z_1z_2=00$ for better exposition. 
By Lemma \ref{ttr}, we obtain $00$ with probability $1/2$ and the other three 2-bits with probability $1/6$. 
Thus, at step 1 we obtain $00$ with probability $a=(1/2)\alpha+(1-\alpha)/4
=1/4+\alpha/4$ and $01,10$ and $11$ with probability $b=(1/6)\alpha+(1-\alpha)/4=1/4-\alpha/12$ for each. 
At step 2, the following four probabilities $q_1,q_2,q_3$ and $q_4$ are important: 
$q_1$ is the probability that $(00,00)$ is obtained; $q_2$ is the probability 
that each of $(00,01),(00,10),(00,11),(01,00),(10,00),(11,00)$ is obtained; 
$q_3$ is the probability that each of $(01,10),(01,11),(10,01),(10,11),(11,01),(11,10)$ is obtained; 
$q_4$ is the probability that each of $(01,01),(10,10),(11,11)$ is obtained. 

$(00,00)$ arises with probability $p_1$ after measuring $00$ and with probability $p_4$ 
after measuring $01,10$ or $11$. We thus have 
\begin{equation}\label{eq1119-1}
q_1= a p_1 +3b p_4
\end{equation}
and similarly
\begin{equation}\label{eq1119-2}
q_2=(a+b)p_2 + 2b p_3,\ \ q_3= 2b p_2 +(a+b)p_3,\ \ q_4=b p_1+(a+2b)p_4.
\end{equation} 
Now let 
\begin{equation}\label{eq1119-3}
q_1=(1/4+\alpha/12)^2,\ q_2=(1/4-\alpha/36)(1/4+\alpha/12),\ \mbox{and}\ q_3=q_4=(1/4-\alpha/36)^2. 
\end{equation}
Then one can easily verify that $q_1+6q_2+6q_3+3q_4=1$. 
Furthermore, the two-qubit state sent to the two outgoing links is 
\begin{align*}
&q_1\pmb{\chi}(00)\otimes\pmb{\chi}(00)+q_2\pmb{\chi}(00)\otimes\pmb{\chi}(01)
+q_2\pmb{\chi}(00)\otimes\pmb{\chi}(10)+q_2\pmb{\chi}(00)\otimes\pmb{\chi}(11)\\
&
+q_2\pmb{\chi}(01)\otimes\pmb{\chi}(00)+q_4\pmb{\chi}(01)\otimes\pmb{\chi}(01)
+q_3\pmb{\chi}(01)\otimes\pmb{\chi}(10)+q_3\pmb{\chi}(01)\otimes\pmb{\chi}(11)\\
&
+q_2\pmb{\chi}(10)\otimes\pmb{\chi}(00)+q_3\pmb{\chi}(10)\otimes\pmb{\chi}(01)
+q_4\pmb{\chi}(10)\otimes\pmb{\chi}(10)+q_3\pmb{\chi}(10)\otimes\pmb{\chi}(11)\\
&
+q_2\pmb{\chi}(11)\otimes\pmb{\chi}(00)+q_3\pmb{\chi}(11)\otimes\pmb{\chi}(01)
+q_3\pmb{\chi}(11)\otimes\pmb{\chi}(10)+q_4\pmb{\chi}(11)\otimes\pmb{\chi}(11),
\end{align*}
which equals 
\[
\left(
\left(\frac{1}{4}+\frac{\alpha}{12}\right)\pmb{\chi}(00)+\left(\frac{1}{4}-\frac{\alpha}{36}\right)
(\pmb{\chi}(01)+\pmb{\chi}(10)+\pmb{\chi}(11))\right)^{\otimes 2}.
\]
Since $\pmb{\chi}(00)+\pmb{\chi}(01)+\pmb{\chi}(10)+\pmb{\chi}(11)=2\pmb{I}$, this can rewritten as
\[\left(
\left(\frac{1}{4}+\frac{\alpha}{12}\right)-\left(\frac{1}{4}-\frac{\alpha}{36}\right)
\right)\pmb{\chi}(00)+\left(\frac{1}{4}-\frac{\alpha}{36}\right)2\pmb{I}
=\frac{\alpha}{9}\pmb{\chi}(00)+\left(1-\frac{\alpha}{9}\right)\frac{\pmb{I}}{2}.
\]
Thus, we obtain the desired two-qubit state. 

What remains to do is to make sure that the values of $p_1,p_2,p_3$ and $p_4$ satisfying  
Eqs.(\ref{eq1119-1}),(\ref{eq1119-2}) and (\ref{eq1119-3}) are all positive and also satisfy Eq.(\ref{eq1111-0}). 
This can be done just by substituting $p_1=\frac{81+6\alpha+\alpha^2}{432}$, 
$p_2=\frac{(9-\alpha)(15+\alpha)}{1296}$, $p_3=\frac{(9-\alpha)(3+\alpha)}{1296}$ 
and $p_4=\frac{9-2\alpha+\alpha^2}{432}$ (all of them are obviously positive for $0<\alpha\leq 1$) 
into Eqs.(\ref{eq1119-1}),(\ref{eq1119-2}),(\ref{eq1119-3}) and (\ref{eq1111-0}). 
Obtaining those values is not so trivial but omitted in this preprint. 
\end{proof}

\subsection{Brief Remarks for the Previous Work}\label{sec:previous-work}
Recall that quantum cloning for a general state \cite{BH96} cannot get rid of a lot of entanglement. 
In \cite{DG98}, Duan and Guo developed a probabilistic cloning system for any fixed two (or more) states, 
which produces, from a given $|\psi\rangle$, state $|\psi\rangle\otimes|\psi\rangle$ with probability $p>0$ 
and an arbitrarily chosen state, say $\frac{\pmb{I}}{2}\otimes\frac{\pmb{I}}{2}$, with probability $1-p$. 
The technique, also based on the fact that the states are fixed, is beautiful but it is quite 
different from ours in the following two senses: (i) Their output state $p\pmb{\psi}\otimes\pmb{\psi}
+ (1-p)\frac{\pmb{I}}{2}\otimes\frac{\pmb{I}}{2}$ is not entanglement-free in the sense of our definition. 
(ii) Their cloning is impossible for any three or more states since they showed that their probabilistic cloning 
can be done if and only if the pure states to be cloned are linearly independent in the sense 
of the vector space of pure state vectors. (Note that the linear independence in Proposition \ref{efc-condition} 
is about the vector space of $2\times 2$ matrices.)

\section{Our Protocol and Its Analysis}

\subsection{Formal Description of the Protocol}\label{1120}

Recall that our current problem is as follows. 

{\bf Input}: A graph $G$ and its CNC protocol $P_C(G)$

{\bf Output}: A QNC protocol $P_Q(G)$ which simulates $P_C(G)$.

We first show a technical lemma about the input graph $G$ and protocol $P_C(G)$. 
A {\em degree-3 (D3)} graph is defined as follows: 
It has five different kinds of nodes, {\em fork nodes}, {\em join nodes}, {\em transform nodes}, 
{\em source nodes}, and {\em sink nodes} whose (indegree, outdegree) is $(1,2)$, $(2,1)$, $(1,1)$, 
$(0,1)$ and $(1,0)$, respectively. 
The classical protocol $P_C(G)$ for a D3 graph is called {\em simple} 
if the operation at each node is restricted as follows: 
(i) The input is sent to the outgoing edge without any change at each source node. 
(ii) The incoming value is just copied and sent to the two outgoing edges at each fork node. 
(iii) The operation of each transform node is constant, one-to-one, or two-to-one. 
(iv) The operation of each join node is the addition (denoted by $+$) 
over $\mathbb{Z}_2\oplus\mathbb{Z}_2$ or $\mathbb{Z}_4$. 
(v) The sink node just receives the incoming value (no operation). 

\begin{lemma}
Without loss of generality we can assume that the input of our problem 
is a pair of a D3 graph and a simple protocol. 
\end{lemma}

\begin{proof}
Assume that a (general) graph $G$ and a protocol $P_C(G)$ are given. 
Then, we transform $G$ and $P_C(G)$ into a 3D graph and a simple protocol as follows. 
If a source node $s$ has $m\geq 2$ inputs, then we add $m$ parent nodes to $s$ as new source nodes 
that have one input for each. Notice that $s$ is no longer a source node. 
Similarly, if a sink node $t$ requires $m\geq 2$ inputs, add $m$ child nodes to $t$ as new sink nodes.
Then, the operations of new sources and sinks clearly satisfy restrictions (i) and (v). 

Next, we decompose nodes of degree $\geq 4$ into fork and join nodes, 
and adapt the classical protocol to the graph changed by the decomposition. 
This is possible since we only consider the operation in the form 
of $\sum_{i=1}^m h_i(X_i)$. For example, Fig.\ 3 is the decomposition of a node such that 
its (indegree, outdegree) is $(3,2)$ and the operations for two outgoing edges are 
$f(X,Y,Z)=a_1X+b_1Y+c_1Z$ and $g(X,Y,Z)=a_2X+b_2Y+c_2Z$, respectively.   
Then, we obtain a D3 graph but we need more transformation to obtain a simple protocol. 
Now the join node has an operation of the form $f(X,Y)=h_1(X)+h_2(Y)$. Recall that 
since our original graph is in $\mathcal{G}_4$, $h_1$ and $h_2$ are constant, two-to-one, 
or one-to-one mapping. We decompose such a join node into two transform nodes $u_1,u_2$ 
and a new join node $u_3$, and design the corresponding protocol as follows: 
$u_1$ and $u_2$ are the parents of $u_3$, the operations of $u_1$, $u_2$ and $u_3$ 
are $h_1$, $h_2$ and $+$, respectively. Then, the new graph is still a D3 graph 
and the new protocol satisfies restrictions (iii) and (iv). What remains to do 
is to satisfy restriction (ii). For this purpose, we delay the operations of 
a fork node, which are written as $h(X)$ for each operation of two outgoing edges, 
until the next transform node (if the next node is a sink, insert an extra transform 
node before the sink). Finally, we have obtained a D3 graph $G'$ and the corresponding 
simple protocol $P_C(G')$. 

We design a quantum protocol for the input $(G',P_C(G'))$ by the algorithm given below. 
Then it is easy to change the protocol back to the protocol for the original graph $G$ 
by combining all the decomposed operations for a node of $G$ into a single operation. 
\end{proof}

\begin{figure}[tb]
\begin{minipage}{0.5\hsize}\begin{center}
\includegraphics*[width=4.5cm]{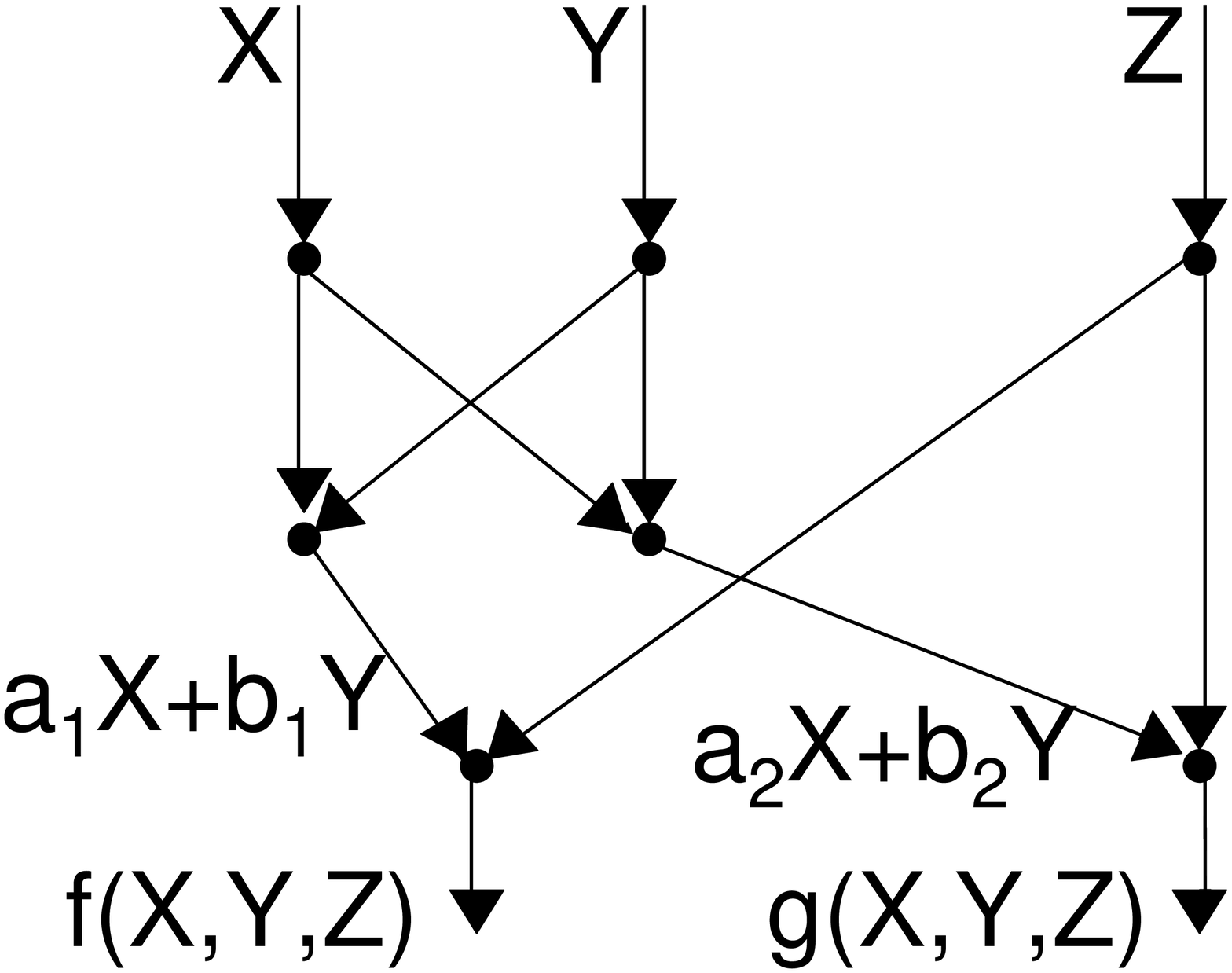}

Figure 3
\end{center}\end{minipage}
\begin{minipage}{0.5\hsize}\begin{center}
\includegraphics*[width=5cm]{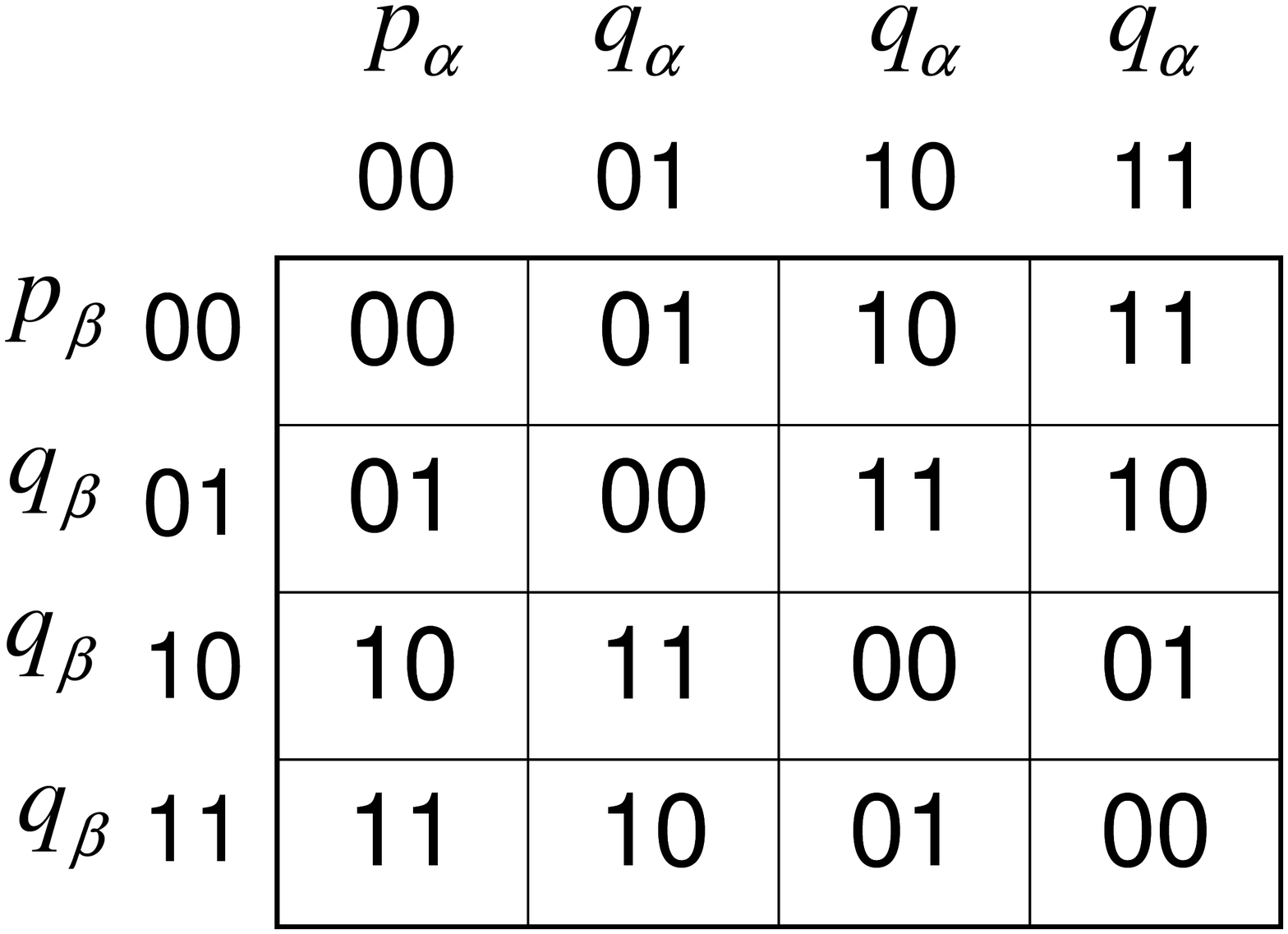}

Figure 4

\end{center}\end{minipage}
\end{figure}

Now we are ready to present our protocol $P_Q(G)$, which is given by the following algorithm 
($Q(v)$ is the operation at a node $v$, and $\alpha(v)$ is the shrinking factor at that node). 

\

{\bf Algorithm for designing $P_Q(G)$.}

{\bf Step 1.} Determine a total order for the vertices of $G$ by their depth ($=$ the length 
of the longest path from a source node). Break ties arbitrarily. Let $v_1,v_2,\ldots,v_r$ be 
their order. 

{\bf Step 2.} For each $v=v_1,v_2,\ldots,v_r$, do the following: 

{\bf If} $v$ is a source node {\bf then} let $\alpha(v)=1$ and let 
$Q(v)=$[Apply $TTR$ for the source, obtain the measurement value $x_1x_2\in\Sigma_4$ 
and send $\pmb{\chi}(x_1x_2)$ to its child node]. 

{\bf Else if} $v$ is a join node {\bf then} let $\alpha(v)=(1/9)\alpha(v_1)\alpha(v_2)$ 
where $v_1$ and $v_2$ are $v$'s parent nodes, and let 
$Q(v)=$[Apply $TTR$ for the two source states, obtain measurement values $x_1x_2\in\Sigma_4$ 
and $y_1y_2\in\Sigma_4$, and send $\pmb{\chi}(x_1x_2+y_1y_2)$ to its child node]. 

{\bf Else if} $v$ is a transform node {\bf then} let $g$ be the corresponding operation 
in $P_C(G)$.

\hspace{1.3cm} {\bf If} $g$ is a constant function, i.e., $g(\cdot)=x_1x_2\in\Sigma_4$ 
{\bf then} let $\alpha(v)=1$ and $Q(v)=$[Send $\pmb{\chi}(x_1x_2)$ to its child].

\hspace{1.3cm} {\bf Else if} $g$ is a one-to-one function {\bf then} let $\alpha(v)=\alpha(v_1)/3$ 
for the parent node $v_1$, and 
$Q(v)=$[Apply $TTR$ for the source state, obtain the measurement value $x_1x_2\in\Sigma_4$ 
and send $\pmb{\chi}(g(x_1x_2))$ to its child].

\hspace{1.3cm} {\bf Else} (i.e., $g$ is a two-to-one function) let $\alpha(v)=\frac{\alpha(v_1)}{6-\alpha(v_1)}$ 
for the parent node $v_1$ and $Q(v)=$[Apply $TTR$ for the source state, obtain the measurement value 
$x_1x_2\in\Sigma_4$, send $\pmb{\chi}(g(x_1x_2))$ to its child with probability $\frac{3}{6-\alpha(v)}$ 
and send $\pmb{\chi}(y_1y_2)$ and $\pmb{\chi}(z_1z_2)$ to its child with probability 
$\frac{3-\alpha(v)}{2(6-\alpha(v))}$ for each, where $\{y_1y_2,z_1z_2\}=\Sigma_4\setminus\mathrm{Range}(g)$].

{\bf Else if} $v$ is a fork node {\bf then} let $\alpha(v)=(1/9)\alpha(v_1)$ for the parent node $v_1$, 
and $Q(v)=$[Apply $EFC_{\alpha(v)}$ for the incoming state and send the resulting two-qubit state 
to its child nodes].

{\bf Else} (i.e., $v$ is a sink node) $Q(v)=$[Do nothing].

\

Our key lemma is as follows. The proof is given in the next subsection.

\begin{lemma}\label{lemma1118}
(i) The value $\alpha(u)$ calculated in the above algorithm is positive for any vertex $u\in V$. 
(ii) Suppose that $P_C(G)$ produces output values $y\in\Sigma_4$ at node $u\in V$ (actually the value 
of the outgoing edge from $u$) from input values $(x_1,\ldots,x_n)\in\Sigma_4^n$. 
Then, if we supply input states $\pmb{\chi}(x_i)$ to source node $s_i$ for $i=1,\ldots,n$, 
then $P_Q(G)$ produces the state $\alpha(u)\pmb{\chi}(y)+(1-\alpha(u))\frac{\pmb{I}}{2}$. 
\end{lemma}

Now we state our main theorem.  

\begin{theorem}
(Main theorem) Suppose that $P_C(G)$ is the same as Lemma \ref{lemma1118} 
and suppose that we supply (general) input states $|\psi_1\rangle,\ldots,|\psi_n\rangle$. 
Then if $P_Q(G)$ produces output states $\pmb{\rho}_1,\ldots,\pmb{\rho}_m$, 
the fidelity between $\pmb{\rho}_i$ and $|\psi_{\sigma(i)}\rangle$ is larger than $1/2$. 
\end{theorem}

\begin{proof}
Let $s$ be the source node that has $|\psi_{\sigma(i)}\rangle$, 
and $t$ be the sink that receives $\pmb{\rho}_i$ by $P_Q(G)$. 
By $TTR$ at $s$ we obtain probabilistic mixture of the four states, 
$\pmb{\rho}=a\pmb{\chi}(00)+b\pmb{\chi}(01)+c\pmb{\chi}(10)+d\pmb{\chi}(11)$. 
By Lemma \ref{lemma1118} (note that the value of $\alpha(u)$ does not depend upon 
the input states $\pmb{\chi}(x_i)$) and linearity we can see that 
$\pmb{\rho}_i= \alpha(t)\pmb{\rho}+(1-\alpha(t))\frac{\pmb{I}}{2}$. 
By Lemma \ref{ttr}, the TP-CP map induced by $TTR$ transforms $|\psi_{\sigma(i)}\rangle$ 
to $\frac{1}{3}|\psi_{\sigma(i)}\rangle\langle\psi_{\sigma(i)}|+\frac{2}{3}\frac{\pmb{I}}{2}$ $(=\pmb{\rho})$. 
Thus, $\pmb{\rho}_i$ is written as $\pmb{\rho}_i=\frac{\alpha(t)}{3}|\psi_{\sigma(i)}\rangle\langle\psi_{\sigma(i)}|
+(1-\frac{\alpha(t)}{3})\frac{\pmb{I}}{2}$. 
Hence we can conclude that the fidelity at $t$ is $\frac{1}{2}+\frac{1}{2}\frac{\alpha(t)}{3}$, 
which is strictly larger than $1/2$. 
\end{proof}

\subsection{Proof of Lemma \ref{lemma1118}}
It is obvious by the algorithm that $\alpha(u)>0$ for all $u\in V$. 
To prove (ii), we need to know what happens at each node. 
We already know the effect of a fork node which is given in Sec.\ \ref{sec:efc}. 
To know the effect of a join node and a transform node, we show two lemmas. 
The first lemma is for a join node.

\begin{lemma}\label{type2}
Assume that $\pmb{\rho}_{x}=\alpha\pmb{\chi}(x_1x_2)+(1-\alpha)\frac{\pmb{I}}{2}$ 
and $\pmb{\rho}_{y}=\beta\pmb{\chi}(y_1y_2)+(1-\beta)\frac{\pmb{I}}{2}$ are sent to a join node. 
Then, the output state is $\frac{1}{9}\alpha\beta\pmb{\chi}(x_1x_2+y_1y_2)
+(1-\frac{1}{9}\alpha\beta)\frac{\pmb{I}}{2}$. 
\end{lemma}

\begin{proof} 
Recall that the operation at a join node is the addition over $\mathbb{Z}_2\oplus\mathbb{Z}_2$ 
or $\mathbb{Z}_4$. Let $f$ be one of such additions. Then one can see that the matrix $M_f=(f(X,Y))$ 
has the property that each value in $\Sigma_4$ appears exactly once in each column and in each row. 
See Fig.\ 4 for the case of $\mathbb{Z}_2\oplus\mathbb{Z}_2$. 
Suppose for example that $x_1x_2=y_1y_2=00$. Then, by Lemma \ref{ttr}, $TTR$ on $\pmb{\rho}_x$ (resp. $\pmb{\rho}_y$) 
produces $00$ (resp. $00$) with $p_\alpha=\alpha/2+(1-\alpha)/4$ 
(resp. $p_\beta=\beta/2+(1-\beta)/4$) and other $01,10$ and $11$ 
with probability $q_\alpha=\alpha/6+(1-\alpha)/4$ (resp. $q_\beta=\beta/6+(1-\beta)/4$) for each. 
Note that $f(00,00)=00$, which appears at four different positions of the matrix whose total probability 
is $r_1=p_\alpha p_\beta+3q_\alpha q_\beta$. Similarly, the value $01$ (similarly for $10$ and $11$) 
appears at four different positions whose total probability is $r_2=p_\alpha q_\beta+p_\beta q_\alpha 
+2p_\beta q_\beta$. By simple calculation, we have $r_1=1/4+\alpha\beta/12$ and $r_2=1/4-\alpha\beta/36$, 
and therefore the output state can be written as 
\[
\left(\frac{1}{4}+\frac{\alpha\beta}{12}\right)
\pmb{\chi}(z_1z_2)+\left(\frac{1}{4}-\frac{\alpha\beta}{36}\right)
(\pmb{\chi}(z_1\bar{z}_2)+\pmb{\chi}(\bar{z}_1z_2)+\pmb{\chi}(\bar{z}_1\bar{z}_2))
=\frac{\alpha\beta}{9}\pmb{\chi}(f(x_1x_2,y_1y_2))+\left(1-\frac{\alpha\beta}{9}\right)
\frac{\pmb{I}}{2}.
\]  
\end{proof}

The second lemma is for the transform node. 

\begin{lemma}\label{change-lemma}
Assume that $\alpha\pmb{\chi}(Z)+(1-\alpha)\frac{\pmb{I}}{2}$ is sent to a transform node 
whose operation in $P_C(G)$ is $g$. 
Then, the output state is $\pmb{\chi}(Z_0)$ if $g$ is a constant function $g(\cdot)=Z_0$, 
$(\alpha/3)\pmb{\chi}(g(Z))+(1-\alpha/3)\frac{\pmb{I}}{2}$ 
if $g$ is one-to-one, and $\frac{\alpha}{6-\alpha}\pmb{\chi}(g(Z))+\left(1-\frac{\alpha}{6-\alpha}\right)
\frac{\pmb{I}}{2}$ if $g$ is two-to-one. 
\end{lemma}

\begin{proof}
The case that $g$ is constant is trivial. The case that $g$ is one-to-one is also easy 
since $TTR$ is the $1/3$-shrinking map (changing the state by $g$ does not lose any fidelity). 
Thus, it suffices to analyze the case that $g$ is two-to-one. 
Assume that $Z'$ is the unique element different from $g(Z)$ in $\mathrm{Range}(g)$. 
(It might help to consider an example such as $g(00)=g(01)=00$, $g(10)=g(11)=10$, $Z=00$ and $Z'=10$.) 
The tetra measurement gives us $Z$ with probability $1/4+\alpha/4$ 
and the other three elements with probability $1/4-\alpha/12$ for each. 
This means that by the calculation of $g$ we obtain $g(Z)$ with probability $\frac{1}{4}+\frac{\alpha}{4}
+\frac{1}{4}-\frac{\alpha}{12}=1/2+\alpha/6$ and $Z'$ with probability $2\times (\frac{1}{4}-\frac{\alpha}{12})
=1/2-\alpha/6$. By our protocol, we obtain $\pmb{\chi}(g(Z))$ 
with probability $(\frac{1}{2}+\frac{\alpha}{6})\frac{3}{6-\alpha}
=\frac{3+\alpha}{2(6-\alpha)}$, $\pmb{\chi}(Z')$ with probability $(\frac{1}{2}-\frac{\alpha}{6})
\frac{3}{6-\alpha}=\frac{3-\alpha}{2(6-\alpha)}$, and the other two tetra states 
$\pmb{\chi}(Y_1)$ and $\pmb{\chi}(Y_2)$ with probability $\frac{3-\alpha}{2(6-\alpha)}$. 
Therefore, the output state, which is their mixed state, is
\[
\frac{3+\alpha}{2(6-\alpha)}\pmb{\chi}(g(Z))+\frac{3-\alpha}{2(6-\alpha)}
(\pmb{\chi}(Z')+\pmb{\chi}(Y_1)+\pmb{\chi}(Y_2))=\frac{\alpha}{6-\alpha}\pmb{\chi}(g(Z))
+\left(1-\frac{\alpha}{6-\alpha}\right)\frac{\pmb{I}}{2}.
\]
This completes the proof.
\end{proof}

Now we prove Lemma \ref{lemma1118} by induction on the depth of nodes. 
First, consider a node $u$ of depth $1$, which has the three cases. 

(Case 1-a: $u$ is a fork node.) 
Let $\pmb{\chi}(x_1x_2)$ be the state sent from a source node $s$. By $EFC_1$ the state 
$\frac{1}{9}\pmb{\chi}(x_1x_2)+\frac{8}{9}\frac{\pmb{I}}{2}$ 
is sent to each of the next two nodes. This clearly satisfies the statement of the lemma 
since $\alpha(u)=(1/9)\alpha(s)=1/9$ (notice that $\alpha(s)=1$ for any source node $s$) 
by the algorithm for designing $P_Q(G)$.

(Case 1-b: $u$ is a join node.) 
Let $\pmb{\chi}(x_1x_2)$ and $\pmb{\chi}(y_1y_2)$ be the states sent from two source nodes 
$s_1$ and $s_2$. By Lemma \ref{type2} we obtain $\frac{1}{9}\pmb{\chi}(x_1x_2+y_1y_2)
+\frac{8}{9}\frac{\pmb{I}}{2}$. This satisfies the statement of the lemma 
since $\alpha(u)=(1/9)\alpha(s_1)\alpha(s_2)=1/9$. 

(Case 1-c: $u$ is a transform node.) 
Let $\pmb{\chi}(x_1x_2)$ be the state received from a source node $s$. By Lemma \ref{change-lemma}, 
we obtain the state $\pmb{\chi}(X_0)$ if the operation $f$ at $u$ in $P_C(G)$ produces 
a constant $X_0\in\Sigma_4$, $(1/3)\pmb{\chi}(x_1x_2)+(2/3)(\pmb{I}/2)$ if $f$ is one-to-one, 
and $(1/5)\pmb{\chi}(x_1x_2)+(4/5)(\pmb{I}/2)$ if $f$ is two-to-one. 
By definition, we can see that $\alpha(u)=1$, $1/3$ and $1/5$ $(=\frac{1}{6-1})$ if $u$ is constant, 
one-to-one, and two-to-one, respectively. Thus, the statement of the lemma holds.

Next, we show that the statement of the lemma holds for any node $u$ at depth $d$ 
under the assumption that it holds for depth $\leq d-1$. 

(Case d-a: $u$ is a fork node.) 
By assumption, $u$ receives a state $\alpha(v)\pmb{\chi}(x_1x_2)+(1-\alpha(v))\frac{\pmb{I}}{2}$ 
from the parent node $v$, where $x_1x_2\in\Sigma_4$ is received at $u$ in $P_C(G)$. 
In the protocol $P_Q(G)$ this state is transformed by $EFC_{\alpha(v)}$. 
By Lemma \ref{EFC-lemma}, the output state is $\left(
\frac{\alpha}{9}\pmb{\chi}(x_1x_2)+(1-\frac{\alpha}{9})\frac{\pmb{I}}{2}\right)^{\otimes 2}$. 
Our algorithm says $\alpha/9=(1/9)\alpha(v)=\alpha(u)$. Thus, the statement of the lemma holds at $u$

(Case d-b: $u$ is a join node.) 
By assumption, $u$ receives two states $\alpha(v_1)\pmb{\chi}(x_1x_2)+(1-\alpha(v_1))\frac{\pmb{I}}{2}$ 
and $\alpha(v_2)\pmb{\chi}(y_1y_2)+(1-\alpha(v_2))\frac{\pmb{I}}{2}$ from the parent nodes $v_1$ 
and $v_2$, where $x_1x_2$ and $y_1y_2$ in $\Sigma_4$ are sent from $v_1$ and $v_2$ to $u$ in $P_C(G)$, 
respectively. Then, by Lemma \ref{type2} the output state is 
$\frac{1}{9}\alpha(v_1)\alpha(v_2)\pmb{\chi}(x_1x_2+y_1y_2)+(1-\frac{1}{9}\alpha(v_1)\alpha(v_2))
\frac{\pmb{I}}{2}$. This satisfies the statement of the lemma 
since $\frac{1}{9}\alpha(v_1)\alpha(v_2) =\alpha(u)$ by our algorithm.  

(Case d-c: $u$ is a transform node.) 
By assumption, $u$ receives a state $\alpha(v)\pmb{\chi}(x_1x_2)+(1-\alpha(v))\frac{\pmb{I}}{2}$ 
from the parent node $v$, where $x_1x_2\in\Sigma_4$ is received at $u$ in $P_C(G)$. 
Let $g$ be the operation at $u$ in $P_C(G)$. Then, by Lemma \ref{change-lemma} 
the output state is $\pmb{\chi}(g(x_1x_2))$ if $g$ is constant, 
$\frac{\alpha(v)}{3}\pmb{\chi}(g(x_1x_2))+(1-\frac{\alpha(v)}{3})\frac{\pmb{I}}{2}$ if $g$ is one-to-one, 
and $\frac{\alpha(v)}{6-\alpha(v)}\pmb{\chi}(g(x_1x_2))+\left(1-\frac{\alpha(v)}{6-\alpha(v)}\right)
\frac{\pmb{I}}{2}$ if $g$ is two-to-one. This satisfies the statement of the lemma since 
for each of the three cases the shrinking factor is $\alpha(u)$ by our algorithm. 

Therefore, by induction we have shown Lemma \ref{lemma1118}.

\section{Concluding Remarks} 
Apparently there remains a lot of future work for EFC.  First of all,
we strongly conjecture that the condition of Proposition \ref{efc-condition} 
is also sufficient. The optimality of our EFC is another interesting research
target.  We also would like to study the opposite direction on the
relation between CNC and QNC, i.e., whether we can derive a CNC
protocol from a QNC protocol.


\appendix

\section{Appendix.\hspace{1cm} Possibility of EFC for Two States}
In this Appendix, we prove that any two pure states (and their shrinking states) admit EFC. 

\begin{proposition}\label{efc-2state}
Let $|\psi_0\rangle$ and $|\psi_1\rangle$ be any different qubits, 
and let ${\cal Q}=\{p|\psi_0\rangle\langle\psi_0|+(1-p)\frac{\pmb{I}}{2}, 
p|\psi_1\rangle\langle\psi_1|+(1-p)\frac{\pmb{I}}{2}\}$ where $p>0$. 
Then ${\cal Q}$ admits EFC.  
\end{proposition}

To prove Proposition \ref{efc-2state}, we first show a lemma which states that 
any two states which are the ``shrinked'' states of 
$|\psi\rangle,|\psi^\bot\rangle$ admit EFC where $|\psi^\bot\rangle$ 
is the orthogonal state to $|\psi\rangle$.  

\begin{lemma}\label{efc-2state-lemma}
Let $|\psi\rangle$ and $|\psi^\bot\rangle$ be any orthogonal qubits. 
The set ${\cal Q}_c=\{p|\psi\rangle\langle\psi|+(1-p)\frac{\pmb{I}}{2}, 
p|\psi^\bot\rangle\langle\psi^\bot|+(1-p)\frac{\pmb{I}}{2}\}$, where $p>0$, admits EFC. 
In fact, there exists an EFC protocol, denoted as $EFCo2_p$, which produces output 
$\left(\frac{p}{2}\pmb{\rho}+(1-\frac{p}{2})\frac{\pmb{I}}{2}\right)^{\otimes 2}$ for a given input  
$\pmb{\rho}\in{\cal Q}_c$.  
\end{lemma}

\begin{proof}
By the symmetry of the Bloch sphere, it suffices to prove the statement for $|\psi\rangle=|0\rangle$ 
and $|\psi^\bot\rangle=|1\rangle$. We then implement the following protocol $EFCo2_p$. 

\

{\bf Protocol $EFCo2_p$.} Input $\pmb{\rho}=p|x\rangle\langle x|+(1-p)\frac{\pmb{I}}{2}$ where $x\in\{0,1\}$. 

{\bf Step 1.} Measure $\pmb{\rho}$ in the basis $\{|0\rangle,|1\rangle\}$, and obtain a bit $X$. 

{\bf Step 2.} Produce the pair $(Y_1,Y_2)$ according to the following probability distribution: 
$(X,X)$ with probability $p_1=1/2+p^2/16$, $(X,\bar{X})$ and $(\bar{X},X)$ with $p_2=1/4-p^2/16$ 
for each, and $(\bar{X},\bar{X})$ with $p_3=p^2/16$.

{\bf Step 3.} Output $|Y_1\rangle$ and $|Y_2\rangle$.

\

After step 2, $EFCo2_p$ produces the pair of bits with the following probability distribution: 
$(x,x)$ with probability $q_1=(1/2+p/2)p_1+(1/2-p/2)p_3=(1/2+p/4)^2$, 
$(x,\bar{x})$ and $(\bar{x},x)$ with probability $q_2=(1/2+p/2)p_2+(1/2-p/2)p_2=(1/2+p/4)(1/2-p/4)$ 
for each, and $(\bar{x},\bar{x})$ with probability $q_3=(1/2+p/2)p_3+(1/2-p/2)p_1=(1/2-p/4)^2$. 
Thus, the final output state is 
\begin{align*}
&q_1|x\rangle\langle x|\otimes |x\rangle\langle x|+q_2(|x\rangle\langle x|\otimes |\bar{x}\rangle\langle \bar{x}|
+|\bar{x}\rangle\langle \bar{x}|\otimes |x\rangle\langle x|)
+q_3|\bar{x}\rangle\langle \bar{x}|\otimes |\bar{x}\rangle\langle \bar{x}|\\
&=\left(\left(\frac{1}{2}+\frac{p}{4}\right)|x\rangle\langle x|+\left(\frac{1}{2}-\frac{p}{4}\right)
|\bar{x}\rangle\langle \bar{x}|\right)^{\otimes 2},
\end{align*}
which equals to $\left((p/2)|x\rangle\langle 
 x|+(1-p/2)\frac{\pmb{I}}{2}\right)^{\otimes 2}$. 
\end{proof}

Using Lemma \ref{efc-2state-lemma} we can prove Proposition \ref{efc-2state}. 

\begin{proofof}{Proposition \ref{efc-2state}}
By symmetry of the Bloch sphere, we prove the statement for $|\psi_0\rangle
=\cos\theta|0\rangle+\sin\theta|1\rangle$ and $|\psi_1\rangle
=\sin\theta|0\rangle+\cos\theta|1\rangle$ where $0\leq\theta < \pi/4$. 
We then implement the following protocol $EFC2_p$. 

\

{\bf Protocol $EFC2_p$.} Input $\pmb{\rho}=p|\psi_x\rangle\langle\psi_x|+(1-p)\frac{\pmb{I}}{2}$. 

{\bf Step 1.} Measure $\pmb{\rho}$ in the basis $\{|0\rangle,|1\rangle\}$, and obtain 
the state $\pmb{\rho}'=p(\cos^2\theta|x\rangle\langle x|+\sin^2\theta|\bar{x}\rangle\langle\bar{x}|)
+(1-p)\frac{\pmb{I}}{2}=p\cos 2\theta|x\rangle\langle x|+(1-p\cos 2\theta)\frac{\pmb{I}}{2}$. 

{\bf Step 2.} Apply $EFCo2_{p\cos 2\theta}$ to $\pmb{\rho}'$, and obtain 
the two-qubit state $\pmb{\rho}''=\left(
\frac{p\cos 2\theta}{2}|x\rangle\langle x|+(1-\frac{p\cos 2\theta}{2})\frac{\pmb{I}}{2}\right)^{\otimes 2}$. 

{\bf Step 3.} For each qubit $\pmb{\sigma}$ of $\pmb{\rho}''$, do the following: 
output $|+\rangle$ with probability $q$ and $\pmb{\sigma}$ with probability $1-q$ 
where $q$ is the positive number determined from $p$ and $\theta$ (seen in the later analysis). 

\

We show that $EFC2_p$ outputs a desired state $(r|\psi_x\rangle\langle\psi_x|+(1-r)\frac{\pmb{I}}{2}
)^{\otimes 2}$ for some $r>0$. It is easy to check that $\pmb{\rho}''$ is obtained at step 2. 
So, we consider what state we obtain after step 3. After step 3, each of two qubits is
\[
\frac{q}{2}\left(\begin{array}{cc}
1 & 1\\
1 & 1
\end{array}
\right)
+(1-q) \left(\begin{array}{ll}
\frac{1}{2}+(-1)^x \frac{p\cos 2\theta}{4} & 0 \\
0 & \frac{1}{2}-(-1)^x \frac{p\cos 2\theta}{4} 
\end{array}
\right),
\]
which should be in the form of $r|\psi_x\rangle\langle\psi_x|+(1-r)\frac{\pmb{I}}{2}$. 
To satisfy this, it suffices that the following equations hold. 
\begin{align*}
& r\cos^2\theta+\frac{1-r}{2}=\left(\frac{1}{2}+\frac{p\cos 2\theta}{4}\right)(1-q)+\frac{q}{2} \\
& r\sin\theta\cos\theta = q/2
\end{align*}
In fact, we can obtain such positive numbers $q$ and $r$ by solving the equations.
This completes the proof. 
\end{proofof}

Furthermore we can show that, for any set ${\cal Q}=\{\pmb{\rho}_1,\pmb{\rho}_2\}$ 
of two mixed state, ${\cal Q}$ admits EFC. Its proof is given by a similar way to 
the proof of Proposition \ref{efc-2state} while we need one extra step as follows: 
(i) By the measurement in a suitable basis, we change the two states $\pmb{\rho}_1,\pmb{\rho}_2$ 
into ``orthogonal'' states $\pmb{\rho}'_1=\alpha|\psi\rangle\langle\psi|+(1-\alpha)\frac{\pmb{I}}{2}$ 
and $\pmb{\rho}'_2=\beta|\psi^\bot\rangle\langle\psi^\bot|
+(1-\beta)\frac{\pmb{I}}{2}$. (ii) If $\alpha\neq\beta$ (say, $\alpha>\beta$), 
change the two orthogonal states to $\gamma|\psi\rangle\langle\psi|
+(1-\gamma)\frac{\pmb{I}}{2}$ and $\gamma|\psi^\bot\rangle\langle\psi^\bot|
+(1-\gamma)\frac{\pmb{I}}{2}$: To do so output the fixed state $|\psi^\bot\rangle$ 
with some probability and the obtained state with the remaining probability. 
(iii) Apply $EFCo2_p$ with a suitable $p$. 
(iv) By outputting $|\psi\rangle$ with some probability, the states can be 
the shrinking states of $\pmb{\rho}'_1$ and $\pmb{\rho}'_2$.  
(iv) Return the angle between the obtained states to that of the original two states $\pmb{\rho}_1,\pmb{\rho}_2$
as step 3 in $EFC2_p$. We can show that this works correctly but omit the verification. 
\end{document}